\documentclass{ifacconf}
\usepackage[hidelinks,implicit=false]{hyperref}
\usepackage{natbib}            % you should have natbib.sty
\usepackage{graphicx}          % Include this line if your
\usepackage{subfigure}

\begin{document}

\begin{frontmatter}

\title{An Introduction to \\Socially Connected Machines: Characteristics and Applications\thanksref{footnoteinfo}} %: \\Characteristics and Applications.}

\thanks[footnoteinfo]{
"This research was supported by the MSIP(Ministry of Science, ICT
and Future Planning), Korea, under the CPRC(Communications Policy
Research Center) support program supervised by the KCA(Korea
Communications Agency)" (KCA-2013-001). }

\author{Taehyoung Shim},
\author{Dong Min Kim} and
\author{Seong-Lyun Kim}

\address{Department of Electrical and Electronic Engineering, Yonsei University\\
50 Yonsei-Ro, Seodaemun-Gu, Seoul 120-749, Korea \\
%\\ 134 Sinchon-Dong, Seodaemun-Gu, Seoul 120 749, Korea
(e-mail: \{teishim, dmkim, slkim\}@ramo.yonsei.ac.kr)}

                             % keyword list or with the
                                          % help of the Automatica
                                          % keyword wizard

\begin{abstract}                          % Abstract of not more than 250 words.

Due to the development of information and communication
technologies, it is difficult to handle the billions of connected
machines. In this paper, to cope with the problem, we introduce
machine social networks, where they freely follow each other and
share common interests with their neighbors. We classify
characteristics and describe required functionalities of socially
connected machines. We also illustrate two examples; a twit-bot and
maze scenario.

\end{abstract}

\end{frontmatter}

%-------------------------------------------%%-------------------------------------------%
%-------------------------------------------%%-------------------------------------------%

\section{Introduction}
Due to the development of information and communication technologies, the number of connections has been increased not only
between people but also between machines \cite{Miorandi:2012}. According to an Erisson report,
there will be 50 billion machines connected to the networks by the end of 2020 \cite{Ericsson:2011}.
In current machine networks, however, it is
difficult to handle
the billions of connected machines \cite{Atzori:2011}.
To manage it more efficiently, it is needed to investigate how the machines
discover the others and connect to one another for exchanging valuable information.

In social network services,
we have seen that a large number of people can be neighbors who have the same interests.
As gathering together in the social networks, people are more close,
$i.e.$ the average number of hops to the others has been decreased.
Thus, they are easy to find their friends who have the same interests \cite{Kwak:2010}.

Through the intuition of social networks,
some previous works changed the point of views on the machine networks.
In \cite{Atzori:2011, Bojic:2012, Java:2007, Kranz:2010}, the authors applied the social concept into the machines
to solve the problem of network scalability.
However, these works still stay showing a what-if scenario.

In this paper, we focus on introducing the characteristics of
machine social networks (MSN), where socially connected machines
(SCMs) freely follow each other and share the same interests with
their followers. In the MSN, SCMs will mediate the interactions not
only between people but also the SCMs, and they carry useful
information to the others.

Throughout this paper, we introduce the characteristics of the SCMs,
and then we suggest social connection elements that are composed of
interest, space, and neighbor axis to assess the characteristics of
the connections. By using the proposed properties, we describe some
examples to illustrate the feasible scenarios in the MSN.

\begin{figure}[!t]\centering
  {\includegraphics[scale=0.38]{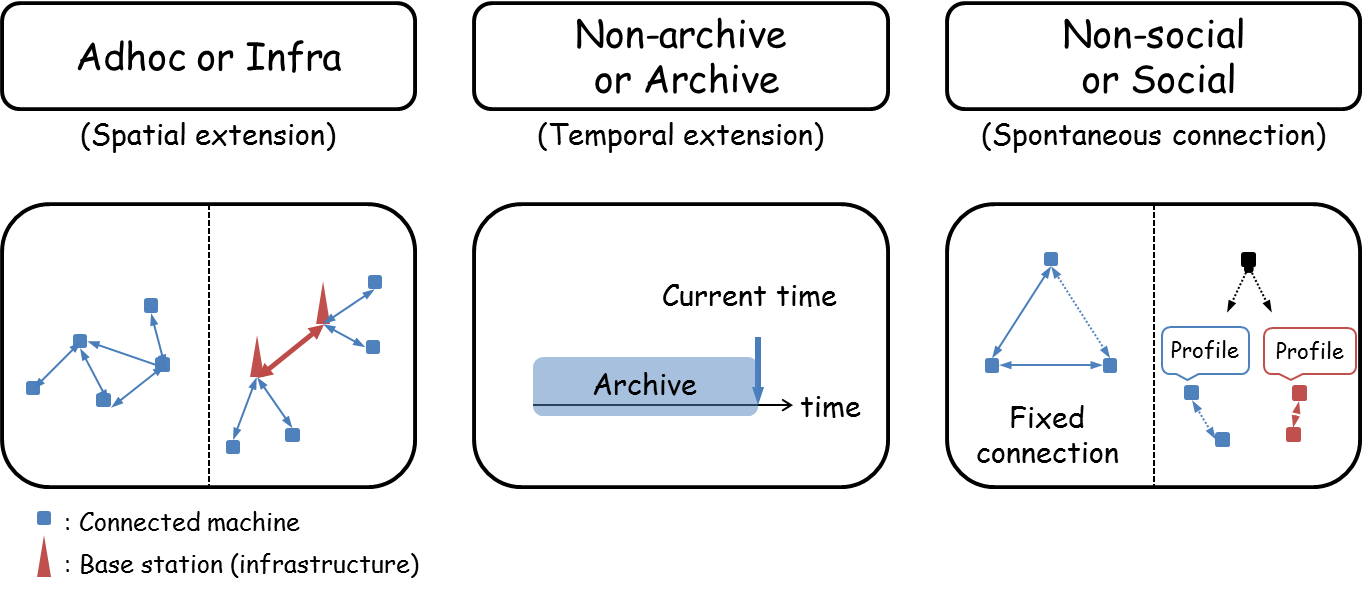}}
  \caption{A taxonomy of machine networks.}\label{fig_taxonomy}
\end{figure}

The rest of this paper is organized as follows:
Section 2 classifies social machine networks.
Section 3 describes the characteristics of the social machines.
In Section 4, we show the examples that indicate feasible scenarios in MSN.
Finally, Section 5 concludes the paper.

%-------------------------------------------%%-------------------------------------------%
%-------------------------------------------%%-------------------------------------------%
\section{Machine networks: Taxonomy}

In this section, we classify machine networks into spatiality,
durability, and sociality as shown in Figure \ref{fig_taxonomy}.
Thus, taxonomy of the machine networks can be explained as below.

\begin{itemize}
  \item {\bf Spatiality}\\
From a spatial perspective, we divide the machines into \emph{infrastructure} and \emph{adhoc} networks.
If the machines use fixed networks (i.e., a wireless access point),
they can access other machines over the limited space by using the Internet.
This case is an infrastructure type network.
On the other hand, in the adhoc type network, the SCMs can only communicate with nearby neighbors,
since the machines communicate via wireless links.
\\
  \item {\bf Durability} %Temporality}\\
Depending on the possibility of accumulation of knowledge,
we can define a durability in the machine networks.
If the machines can freely use storages that are shared with the others,
then this case is an \emph{archive} type network.
On the contrary, the network of the machines is a \emph{non-archive} type if they can share only volatile information.
\\
  \item {\bf Sociality}\\
The machines share the information with the others by using the
communication capability. Depending on the degree of a connection
freedom, we divide the type of the machine networks into
\emph{non-social} network and \emph{social} network. The machine
network is non-social if the machine in the network is controlled by
people or can share the information with only limited machines. The
connection is fixed in the non-social network. On the other hand, if
the connected machines freely follow each other and spontaneously
share the information with their followers, the network is social.
\end{itemize}

According to the taxonomy of the machine networks, the machines belong to each of the classifications.
The machines using social connection are called the socially connected machines (SCMs).
In the next section, we describe the SCMs in detail.

%-------------------------------------------%%-------------------------------------------%
%-------------------------------------------%%-------------------------------------------%

\section{Socially Connected Machines}
For achieving purposes, the SCMs
connect the others with the same interests spontaneously to cooperate mutually.
The SCMs have high connection diversity, public information and the ability to have multiple purposes. To realize the MSNs, SCMs should have following functionalities:
announcing identification, discovering neighbors, having multiple interests and determining trustworthiness.

%-------------------------------------------%%-------------------------------------------%
\subsection{Characteristics}

\begin{figure}[!t]\centering
  {\includegraphics[scale=0.5]{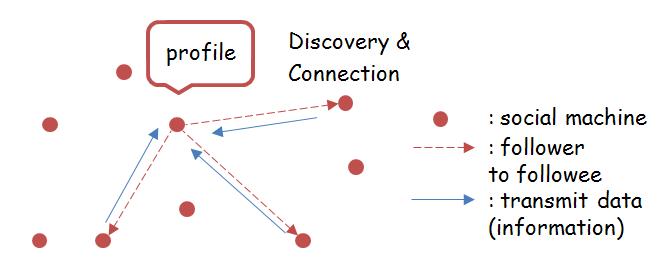}}
  \caption{Description of machine social networks. %: How social machines connect to each other.
}\label{fig_socialmachine}
\end{figure}

In Figure \ref{fig_socialmachine}, we illustrate the structure of
the MSNs. The machines who want to obtain information from their
neighbors compose a network. The information provider is followee,
and the receivers are followers. To discover the appropriate
information provider, each machine investigates the profiles of the
neighbors. By using their profiles, the SCMs receive the information
from them. The characteristics of the SCMs are explained as below.

\begin{itemize}
  \item {\bf Connection diversity}\\
One of the characteristics is connection diversity.
As mentioned in the previous section,
the non-social machines have the fixed connection that only share the piece of information
with limited machines.
On the other hand, the SCMs have a variety of the connections depending on the interests.
The possible number of connections between SCMs is greater than the number of connections in non-social network.
As varying the space and time,
the SCMs choose various machines to share the interesting information.
Therefore, the diversity of the connections increases.
In the MSN, the connections changes more frequently than non-social network.
\\
  \item {\bf Public information}\\
Each machine should open to the public the profile to allow
accessing it. In MSNs, any SCM can freely access and widely
disseminate the information to the others. Because of the lower
privacy standards for the machines, the information propagation is
relatively easy. The SCMs may be vulnerable from a spread of
misinformation and a malicious attack.
\\
  \item {\bf Multiple purposes}\\
The SCMs have not only one-purpose but multi-purpose.
Since their interests vary in different space and time,
they can serve multiple applications.
As allowing multi-purpose to the machines, we reduce cost of machines efficiently.
\\
\end{itemize}

\begin{figure}[!t]\centering
  {\includegraphics[scale=0.55]{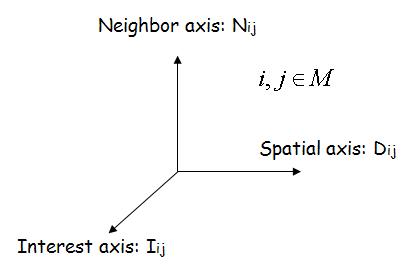}}
  \caption{Interest, spatial, and neighbor axis in machine social networks.}\label{fig_socialaxis}
\end{figure}

%-------------------------------------------%%-------------------------------------------%
\subsection{Required Functionalities}
To implement the SCMs, following functionalities are required.

\begin{itemize}
  \item {\bf Identification}\\
The SCMs need a machine profile
that describes their identification information, such as operating time, location, interests, capabilities (e.g., kinds of sensors, peak velocity, amount of memory).
\\
  \item {\bf Discovery neighbors}\\
The SCMs should autonomously discover their neighbors who have the same interests.
\\
  \item {\bf Multiple interests}\\
The SCMs can have the multiple interests and
a capability of distributed tasks.
\\
  \item {\bf Trustworthiness}\\    %%Reliable credit}
The SCMs should have a capability that determines the reliability of other machines \cite{Atzori:2011}.
If the SCM connects with the malicious machine and is attacked, the SCM should exclude the spiteful machines from MSN.
\end{itemize}

%-------------------------------------------%%-------------------------------------------%
\subsection{Connection between SCMs}\label{section:connection}

In this subsection,
we explain that how the SCMs find their neighbors who have the same interests.
To make possible the social network of machines, they have to discover the neighbors by using their profiles.

%-------------------------------------------%%-------------------------------------------%

A relationship between machines $i$ and $j$ is established when the condition of the connected machines is met.
Let $C_{ij}$ be the measure of the strength of the connection from $i$ to $j$.
If the following condition is satisfied, the connection is established:
\begin{equation}\label{eq_best_response}
  C_{i j} \geq C_{th},~ i, j \in M,
\end{equation}
where $C_{th}$ is the threshold to make a connection between machines, and $M$ denotes a set of the machines in machine networks.
The threshold means a capability of machines in order to communicate each other.
The social machine $i$ follows the machine $j$ if the relationship between $i$ and $j$ is bigger than the threshold.
For discovering, we consider three elements: interest, spatial, neighbors factor between two machines as shown in Figure \ref{fig_socialaxis}.
We define the relation between the machine $i$ and $j$ as

\begin{equation}\label{eq_connection}
  C_{i j} = w_{I}I_{i j} + w_{D}D_{i j} + w_{N}N_{i j},
\end{equation}

\begin{figure}[!t]\centering
  {\includegraphics[scale=0.28]{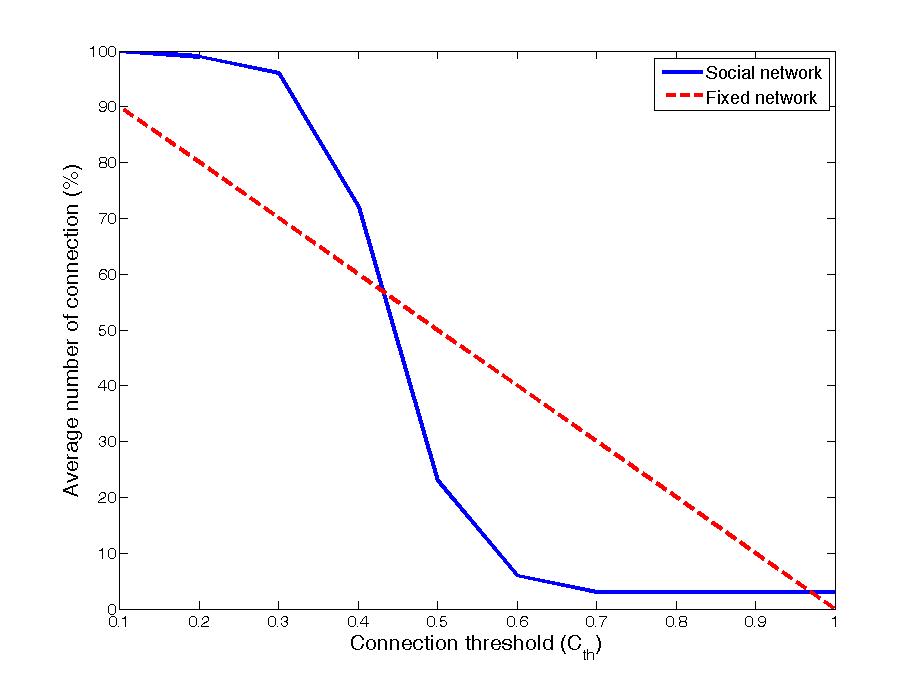}}
  \caption{Average number of connections.}\label{fig_simulation}
\end{figure}

where weighted factors of the relationships are
$w_{I}+w_{D}+w_{N}=1$. In (\ref{eq_connection}), first, the interest
axis is $I_{i j} = |I(i) \cap I(j)|/|I(i)|$, where $|I(i)|$ denotes
the number of interests of the machine $i$. It means a correlation
of interests between the machine $i$ and $j$ in the machine
networks. If the machine $i$ and $j$ are working simultaneously with
all of the same interests, a value of the correlation is equal to
one. Second, the spatial axis, $D$ represents $D_{i j} = 1 - dist
(x(i) - x(j))/|x|$, where $x(i)$ means location of the machine $i$,
$dist(\cdot)$ is the Euclidean distance. It means distance between
two machines. The maximum distance between two machines is $|x|$. As
close as possible, the spatial correlation between two machines
$D_{i j}$ goes to one. Finally the social axis, $S$ means that how
many they have mutual neighbors. The mutual neighbors is $N_{i j} =
|N(i) \cap N(j)|/|N(i)|$, where $N(i)$ denotes the set of neighbors
of machine $i$. If two machines have a lot of mutual neighbors, then
they make a neighbor relationship more easily.

As time goes by, the connected machines have the same level of information. It means the follower has no reason to follow the followee.
We model this situation as follows:
\begin{equation}
  C_{ij}(\Delta t) = e^{- a \Delta t} < C_{th},
\end{equation}

where $a$ is a constant value, and $\Delta t$ denotes an elapsed time from a time of making a neighbor. The strength of the social relation between two machines decays exponentially.
In the MSNs, the social connection between two machines makes or deletes a link repeatedly.

\begin{figure}[!t]\centering
    \includegraphics[width=3.3in]{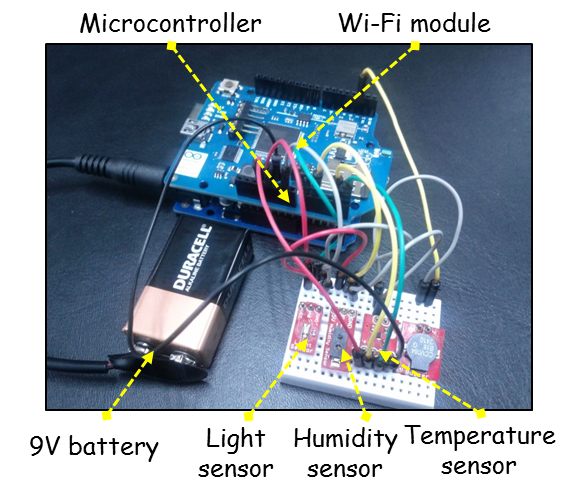}
    \label{fig:twitbot_testbed}
    \caption{A socially connected machine with sensors.}\label{fig_twitbot}
\end{figure}

%-------------------------------------------%%-------------------------------------------%
\subsection{Social Connection Simulation}
To evaluate the performance of the SCMs, we simulate the MSNs and show that how the average number of connections changes with various connection thresholds. %C_{th}.
To compare to the performance, we use non-social networks with fixed
connections.

We set the total number of machines is 100.
In one dimensional space, there are 10 subspaces.
Each machine exists one of subspaces. If there are two machines in the same subspace,
then the distance is zero and $D_{ij}$ is one.
The total number of interests is ten, and each machine has five interests.
In this network, the SCMs make neighbors and disconnect with them repeatedly.
As we mentioned the threshold $C_{th}$, it means a capability of machines in order for communicating each other.
If it goes to one, then making the connection between the machines is very low.
If the threshold is extremely high, the machines can not make a group and be separated in the network.

%-------------------------------------------%%-------------------------------------------%
\subsubsection{Simulation result}

In Figure \ref{fig_simulation}, we evaluated the average number of connection according to the connection threshold $C_{th}$.
The connection threshold is a technology that is correlated with discovering other machines and transmitting capability.
If the connection threshold is low, then a dissemination of information is high.
As shown in Figure \ref{fig_simulation}, the SCMs (solid line) is isolated if the connection threshold is low.
It is shown that the number of connection between the machines is higher than typical fixed machine networks (dotted line)
if the connection threshold is lower than 0.45.
This means that the SCMs can make a connection or a group if the machines have lower threshold.
This simulation result shows that the connection threshold should be relatively low
in order to take advantage of the MSN.

%-------------------------------------------%%-------------------------------------------%
%-------------------------------------------%%-------------------------------------------%
\section{Machine Social Network Applications}
%\section{Examples}

In this section, we suggest two examples in MSN.
One is a twit robot, and the other is a maze scenario.
According to the machine taxonomy, the twit bot has properties; infrastructure, archive, and social.
Another example has properties; Adhoc, Non-archive or Archive, and Social.

%-------------------------------------------%%-------------------------------------------%

%\section{Social Machine Testbed}\label{section:testbed}
\subsection{Twitbot}

The SCMs transmit the information to the followers automatically. They are able to discover neighbors who have the same interests and connect with them.

As shown in Figure \ref{fig_twitbot}, we installed one example of the SCMs.
Our machine is composed of a micro controller, a wireless module, and several sensors.
We use $arduino~uno$ board that is easy to be controlled a micro controller by using c language.
In order to connect to the Internet conveniently, we adopted $arduino$ $wi$-$fi$ $shield$ that conforms to standard IEEE 802.11 b/g  \cite{Arduino:2013}.
Our machine also has sensors; light, humidity, and temperature sensors. From these sensors, we could sense our environments.

This machine periodically perceived changes in surrounding environment, for example, notifying weather changes, and posting critical information on its web page,
such as $Twitter$.
Followers of this social machine can receive the information periodically or in emergency situations.
Our SCM periodically sends the information to its followers.
In order to make a social relationship with others, it should discover others who have the same interests.
%In the next section, we design a social connection algorithm between the social machines.
%social relationship between follower and followee are different depending on the interests.

\begin{figure}[!t]\centering
    \includegraphics[width=2.5in]{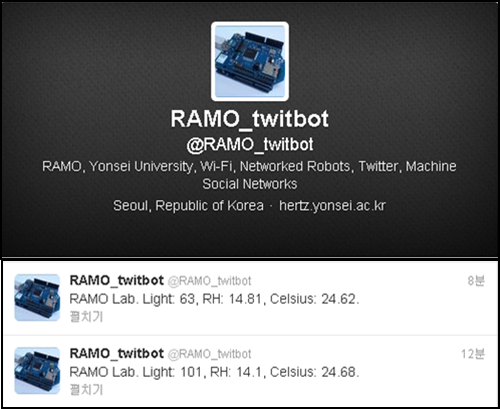}
    \label{fig:maze_archive}
    \caption{Social network platform: Twitter.}
\end{figure}

%-------------------------------------------%%-------------------------------------------%

\subsection{Socially Connected Machines in Maze}

In \cite{Jung:2010}, the authors referred that
machines can cooperatively solve a maze problem by using wireless multi-hop communications.
They utilized the multi-hop routing protocol that is proposed by \cite{Kim:2011, Shim:2012}.
The maze is a tour puzzle that is composed of roads and walls.
From the maze scenario, we describe a storage of knowledge in maze by using a cooperation between SCMs.

\begin{figure}[!t]\centering
  {\includegraphics[scale=0.6]{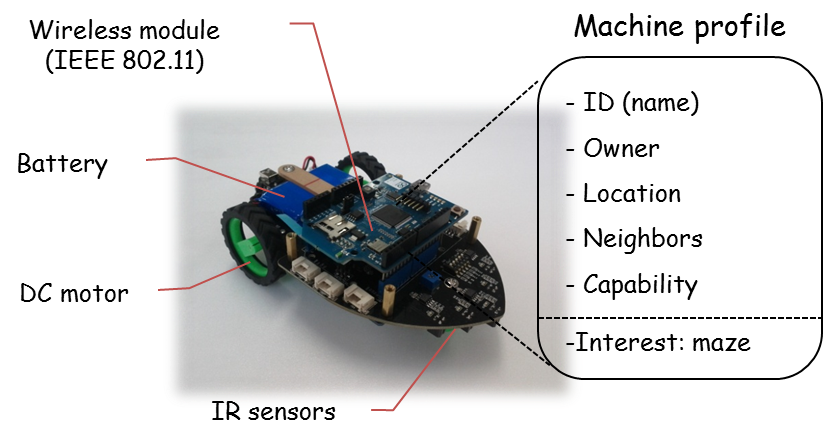}}
  \caption{Socially connected machine in a maze.}\label{fig_mazemachine}
\end{figure}

\subsubsection{Social machine testbed}
As shown in Figure \ref{fig_mazemachine}, we set up an SCM in the maze scenario.
It is composed of sensors (detecting roads), actuators (DC motor), a battery (3V),
and a wireless module (IEEE 802.11).
In a memory of the machine, it remembers its profile; identification, owner, location information,
its neighbors, capabilities, and current interests.
By using the wireless module, the machine periodically transmits the machine profile to the other machines.

\begin{figure}
    \centering
    \subfigure[Discovery neighbor.]
    {
        \includegraphics[width=3.7in]{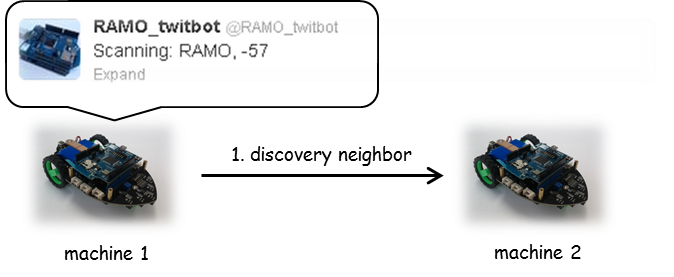}
        \label{fig:maze_discover}
    }
    \\
    \subfigure[Connect and share informations.]
    {
        \includegraphics[width=3.7in]{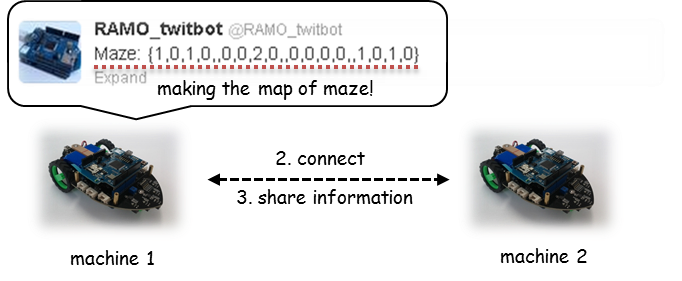}
        \label{fig:maze_connect}
    }
    \\
    \subfigure[Access to archive.]
    {
        \includegraphics[width=3.3in]{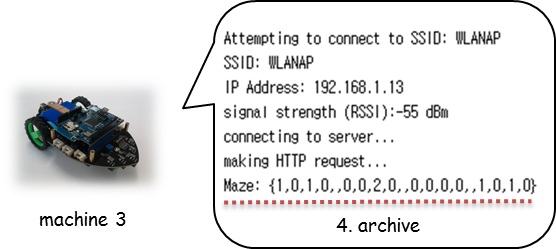}
        \label{fig:maze_archive}
    }
    \caption{Machine social networks scenario in maze.}
    \label{fig:social_machine_maze}
\end{figure}

\subsubsection{Maze scenario}
In Figure \ref{fig:social_machine_maze}, we make a description of MSN scenario in the maze,
where the social machines try to find a route.
There are two social machines in the maze, as shown in Figure \ref{fig:maze_discover}.
From broadcasting their profiles, each of machines try to discover the other that has the same interest, the maze.
If they find each other, and then they share the map information in Figure \ref{fig:maze_connect}.
Here, we use a chat server, in which the SCMs can share their map information with other machines.
By using the completed map information, the SCMs find their route in the maze.
If one of the SCMs connect to a storage, for example a cloud server,
then it sends the information to the storage.

After escaping, another SCM, a machine 3, enters to the maze.
As shown in Figure \ref{fig:maze_archive}, the SCM accesses to archive and find the map information
that is completed by the machine 1 and 2.
If the machine 3 uses the archive, then it can be more easily find the exit in the maze.

%-------------------------------------------%%-------------------------------------------%
%-------------------------------------------%%-------------------------------------------%

\section{Discussion and Future Work}\label{section:discussion}
In this paper, we have suggested the socially connected machines to
manage the machines more efficiently. To cope with the scalability
problem, we described the characteristics and the requirements
capabilities for the socially connected machines. Then, we showed
two examples in the machine social networks; the twitbot and the
maze machines. As part of the future work, we will set up a testbed
for the scenario. We plan to additionally include a socially connect
algorithm to connect between follower and followee. Also, we
consider that solve a malicious or a broken machine problem.

\end{document}